\documentclass{emulateapj}
\slugcomment{}

\usepackage{pslatex}
\usepackage{setspace}

\def\pA{498.9}
\def\peA{1.0}
\def\pyearsA{1.366}
\def\pyearseA{0.003}
\def\tpA{2453057}
\def\tpeA{5}
\def\eA{0.71}
\def\eeA{0.04}
\def\omA{285}
\def\omeA{7}
\def\kA{20.8}
\def\keA{1.5}

\def\msiniA{0.68}
\def\msinieA{0.18}
\def\arelA{1.3}
\def\areleA{0.07}
\def\rmsA{4.6}
\def\chisA{0.88}
\def\nobsA{47}

\def\mstarA{1.31}
\def\mstareA{0.09}
\def\bvA{0.73}
\def\vmagA{8.09}
\def\mvA{3.46}
\def\vsiniA{3.8}
\def\vsinieA{0.5}
\def\ageA{3.8}
\def\rstarA{1.86}
\def\rstareA{0.07}
\def\lstarA{3.4}
\def\lstareA{0.2}
\def\bcorrA{-0.081}
\def\teffA{5770}
\def\teffeA{70}
\def\loggA{4.0}
\def\loggeA{0.2}
\def\feA{$+0.34$}
\def\mfeA{+0.34}
\def\feeA{0.06}

\def\dA{84}
\def\deA{9}

\def\pB{990}
\def\peB{20}
\def\pyearsB{2.71}
\def\pyearseB{0.05}
\def\tpB{2452820}
\def\tpeB{30}
\def\eB{0.59}
\def\eeB{0.11}
\def\omB{222}
\def\omeB{9}
\def\kB{94}
\def\keB{11}

\def\msiniB{4.4}
\def\msinieB{0.34}
\def\arelB{2.1}
\def\areleB{0.08}
\def\rmsB{9.2}
\def\chisB{1.16}
\def\nobsB{44}

\def\mstarB{1.29}
\def\mstareB{0.09}
\def\bvB{0.63}
\def\vmagB{7.29}
\def\mvB{3.40}
\def\vsiniB{4.8}
\def\vsinieB{0.5}
\def\ageB{4.0}
\def\rstarB{1.68}
\def\rstareB{0.04}
\def\lstarB{3.49}
\def\lstareB{0.12}
\def\bcorrB{-0.035}
\def\teffB{6080}
\def\teffeB{70}
\def\loggB{4.4}
\def\loggeB{0.2}
\def\feB{$+0.39$}
\def\mfeB{+0.39}
\def\feeB{0.06}

\def\dB{60}
\def\deB{3}

\newcommand{\starA}{HD\,96167}
\newcommand{\starB}{HD\,16175}

\newcommand{\shkA}{0.14} 
\newcommand{\rhkA}{$-5.16$}
\newcommand{\shkB}{0.17}  
\newcommand{\rhkB}{$-4.96$}
\newcommand{\hip}{{\em Hipparcos}}

\newcommand{\lsps}{Lick Subgiants Planet Search}
\newcommand{\mA}{\textrm{A}}
\newcommand{\mB}{\textrm{B}}
\newcommand{\mC}{\textrm{C}}
\newcommand{\mT}{\textrm{T}}
\def\arraystretch{1.200}

\newcommand{\unit}[1]{\textrm{ #1}}

\newcommand{\ms}{m$\:$s$^{-1}$}
\newcommand{\lsun}{L$_{\odot}$}
\newcommand{\msun}{M$_{\odot}$}
\newcommand{\rsun}{R$_{\odot}$}

\newcommand{\kms}{km$\:$s$^{-1}$}
\newcommand{\teff}{$T_{\rm eff}$}
\newcommand{\mteff}{T_{\rm eff}}
\newcommand{\logg}{$\log{g}$}
\newcommand{\feh}{[Fe/H]}
\newcommand{\caii}{Ca~\textsc{ii}}
\newcommand{\msini}{M_{P}\sin{i}}

\newcommand{\vsini}{$v\sin{i}$}
\newcommand{\mjup}{M$_{\rm Jup}$}
\newcommand{\chisq}{$\sqrt{\chi^2_{\nu}}$}

\newcommand{\sA}{\,\AA}

\received{2009 March 20}
\accepted{2009 April 16}
\published{in PASP, 2009 May 16}
\shortauthors{Peek et~al.\ }
\shorttitle{Two Eccentric Jovian Planets}

\begin{document}

\title{Old, Rich, and Eccentric: Two Jovian Planets Orbiting Evolved Metal-Rich Stars\altaffilmark{1}}
\author{Kathryn M.~G.~Peek,\altaffilmark{2}
        John Asher Johnson,\altaffilmark{3}
        Debra A.~Fischer,\altaffilmark{4}
        Geoffrey W.~Marcy,\altaffilmark{2}
        Gregory W.~Henry,\altaffilmark{5}
        Andrew W.~Howard,\altaffilmark{2}
        Jason T.~Wright,\altaffilmark{6}
        Thomas B.~Lowe,\altaffilmark{7}
        Sabine Reffert,\altaffilmark{8}
        Christian Schwab,\altaffilmark{8}
        Peter K.~G.~Williams,\altaffilmark{2}
        Howard Isaacson,\altaffilmark{4}
        Matthew J.~Giguere\altaffilmark{4}
}
\email{kpeek@astron.berkeley.edu}

\altaffiltext{1}{Based on observations obtained at the Lick Observatory, which is operated by the University of California, and on observations obtained at the W.\,M.\ Keck Observatory,
which is operated as a scientific partnership with the University of California, the California Institute of
Technology, and the National Aeronautics and Space Administration. Keck Observatory was made possible by the generous financial support of the W.\,M.\ Keck Foundation. Keck time has been granted by both NASA and the University of California.}
\altaffiltext{2}{Department of Astronomy, University of California, 
Berkeley, CA, 94720-3411.}
\altaffiltext{3}{Institute for Astronomy, University of Hawaii, Honolulu, HI, 96822.}
\altaffiltext{4}{Department of Physics and Astronomy, San Francisco State University,
San Francisco, CA, 94132.}
\altaffiltext{5}{Center of Excellence in Information Systems,
Tennessee State University, 3500 John A.~Merritt Blvd., Box 9501, Nashville, TN 37209, USA}
\altaffiltext{6}{Department of Astronomy, 610 Space Sciences Building, Cornell University, Ithaca, NY, 14853}
\altaffiltext{7}{UCO/Lick Observatory, 1156 High Street, Santa Cruz, CA, 95064}
\altaffiltext{8}{ZAH-Landessternwarte, K\"onigstuhl 12, 69117 Heidelberg, Germany}

\begin{abstract}
We present radial velocity measurements of two
stars observed as part of the \lsps\ and the Keck
N2K survey.
Variations in the radial velocities of both stars
reveal the presence of Jupiter-mass 
exoplanets in highly eccentric orbits.
\starB\ is a G0 subgiant from the \lsps,
orbited by a planet having a minimum mass of 
$\msiniB\unit{\mjup}$, in an eccentric 
($e = \eB$), $\pyearsB\unit{yr}$ orbit. 
\starA\ is a G5 subgiant from the N2K (``Next 2000'') 
program at Keck Observatory, orbited by a 
planet having a minimum mass of 
$\msiniA\unit{\mjup}$, in an eccentric 
($e = \eA$), $\pyearsA\unit{yr}$ orbit.
Both stars are relatively massive
($M_{\star}=1.3\unit{\msun}$) and are very metal rich
(\feh~$ > +0.3$).
We describe our methods for measuring the stars' radial
velocity variations and photometric stability.
\end{abstract}


\section{Introduction}
\label{introduction}
Over 300 exoplanets have been detected to 
date,\footnote{\texttt{http://www.exoplanets.org/}} 
a quantity that is making it possible
to characterize statistically significant trends that 
shed light on planet formation processes.
Trends in planet occurrence as a function of
host star mass and metallicity have
emerged, and the distribution of orbital
eccentricities is still being uncovered.
For a review of exoplanet detections, see, e.g.,
\citet{johnson_2009,marcy_2008,udry_2007}.

Both the planet occurrence rate and 
the total planetary mass in a system are proving to be
positively correlated with the host star's mass
\citep{johnson_2007a,lovis_2007}.
Our understanding of that planet occurrence rate at masses 
of $M_{\star} \geq 1.3\unit{\msun}$ comes from surveys of 
evolved stars.
Intermediate-mass stars cool and slow their 
rotation as they evolve off the main sequence, thereby 
developing the copious narrow metal lines that facilitate 
precision radial velocity monitoring.
Since giant stars tend to have considerable photospheric 
jitter (introducing errors of $\sim$\,$20\unit{\ms}$),
subgiants, which have jitters of $\sim$\,$5\unit{\ms}$,
represent an evolutionary sweet spot for detecting 
planets around intermediate-mass stars.

In addition to correlating with stellar mass, 
the planet occurrence rate is observed to correlate with stellar
metallicity \citep{fischer_2005,santos_2004,gonzalez_1997}.
The results of Doppler surveys are
corroborated by theoretical predictions that more metals 
in a protoplanetary disk leads to more prolific planet
formation \citep{ida_2004}.
This trend informs target selection for some exoplanet searches, 
e.g., the N2K program \citep[``Next 2000,''][]{fischer_2005n2k}
and the ELODIE Metallicity-Biased Planet Search \citep{dasilva_2006}, 
which specifically target metal-rich stars. 
Such searches allow planet hunters to 
maximize exoplanet discoveries given finite
telescope resources.

Exoplanet systems exhibit a broad range of eccentricities.
For the 267 well-characterized planets 
within $200\unit{pc}$ (most of the known exoplanets), 
those in non-tidally-circularized 
orbits ($P > 5\unit{days}$) span a range of eccentricities 
from 0.00 to 0.93, with a median eccentricity of 0.23.
The broad range of eccentricities is surprising, 
since it was previously expected that such
objects would have orbits resembling the circular orbits
of the solar system gas giants.
The origin of exoplanet eccentricities is one of the major
outstanding problems in the study of exoplanets.
The discovery of additional planetary systems, especially 
those with large eccentricity, will bring the shape 
of the eccentricity distribution into better focus and
guide theoretical models of planet formation and orbit
evolution \citep[e.g.,][]{ford_2008,juric_2008,chatterjee_2008}.
                                   
We present here two new Jupiter-mass exoplanets detected with 
the radial velocity method.
Both stars are metal-rich (\feh~$ > +0.3$),
intermediate-mass ($M_{\star} = 1.3\unit{\msun}$),
G-type subgiants, and therefore contribute to our 
understanding of the planet occurrence rate as a 
function of mass and metallicity.
\starB\ is part of the \lsps\ \citep{johnson_2006} and
\starA\ is part of the N2K program at Keck Observatory
\citep{fischer_2005n2k}.
Because the planets presented here are from different
programs at different observatories, we describe the 
observational, data analysis, and Keplerian techniques 
for each star in its own section. 
\starB\ is presented in \S\,\ref{star_b} and \starA\ in 
\S\,\ref{star_a}. 
We review the photometric stability of the stars
in \S\,\ref{phot} and summarize our results in
\S\,\ref{summary}.

\section{\starB}    
\label{star_b}

\subsection{Observations and Radial Velocity Measurements}
\label{rv_measurements_b}

We began Doppler monitoring of \starB\ (HIP\,12191, BD+41\,496) 
at Lick Observatory in 2004 November as part of the \lsps,
a program comprising 159 evolved
intermediate-mass stars, chosen
to have color index $0.55 < B - V < 1.00\unit{mag}$, 
$V$-magnitude brighter than $7.6\unit{mag}$, 
and declination $d > -20^{\circ}$.
In order to choose evolved stars, we require an absolute visual
magnitude $M_{V}$ more than 1.0 mag above the main sequence.
For this purpose, we use the main sequence defined in
\citealt{wright_2005}, which is a ninth-order polynomial
fit to the \hip\ dwarf stars within $60\unit{pc}$. 
In order to
ensure selection of subgiants rather than giants, we required
$M_{V}$ fainter than $1.0\unit{mag}$. 
The color limits select stars for which radial velocity monitoring
is feasible, while the $V$-magnitude and declination limits allow 
the sample to be observed with the \mbox{3 m} Shane telescope and 
with the \mbox{0.6 m} Coud\'e Auxiliary Telescope (CAT), both of 
which feed the Lick Observatory Hamilton echelle spectrograph 
\citep{vogt_1987}. 
See \citet{johnson_2006} for a complete
description of the \lsps\ and its selection criteria.

We use the Hamilton spectrograph at a setting that provides
resolution $R \approx 50\,000$ at $\lambda = 5500$\sA.
On the CAT, typical exposure times were 60 minutes, yielding 
signal to noise ratio (SNR) of $\sim$\,$50$ at 5500\sA. 
On the Shane, shorter 10-minute 
exposures generally yielded higher SNR~$\sim 175$. 
The lion's share of the data presented here 
were collected with the CAT.
The procedure for obtaining radial velocities from Hamilton 
spectra is outlined in \citet{butler_1996}. 
In short, a temperature-controlled Pyrex cell, filled with 
gaseous molecular iodine, is placed before the spectrometer's 
entrance slit. 
The iodine imprints a dense set of absorption lines between 
$\lambda \sim 5000$\sA\ and $\lambda \sim 6400$\sA, 
which are calibrated to provide an accurate wavelength scale for 
each observation, and which also serve to provide information 
about the spectrometer's instrumental response \citep{marcy_1992}.

The Doppler measurement method described above traditionally
includes the use of a template
spectrum, an iodine-free, high-S/N observation 
relative to which the Doppler radial velocity measurements are 
made. 
In the case of \starB, we at first measured the Doppler shifts
relative to a synthetic ``morphed'' template spectrum.
The procedure for creating the morphed template is described
in \citet{johnson_2006}. 
Once the star showed the potential for a planetary companion, 
we obtained a traditional template observation. 
The observed template is also used for the LTE analysis described in 
\S\,\ref{stellar_properties_b}.

\def\arraystretch{1.100}
\begin{deluxetable}{rrr}
\tablecaption{Radial Velocities for HD\,16175.\label{vel16175}}
\tablewidth{0pt}
\tablehead{
 &
 \colhead{Radial Velocity} &
 \colhead{Uncertainty} \\
\colhead{\makebox[3cm][c]{${\rm JD} - {\rm 2,400,000}$}} &
 \colhead{(\ms)} &
 \colhead{(\ms)} 
}
\startdata
$13338.681$ \dotfill &    $81.17$ &  5.96 \\
$13599.929$ \dotfill &    $19.72$ &  4.64 \\
$13629.867$ \dotfill &     $5.40$ &  4.73 \\
$13668.833$ \dotfill &   $-10.57$ &  4.96 \\
$13710.744$ \dotfill &   $-48.24$ &  5.36 \\
$13718.692$ \dotfill &   $-46.62$ &  5.26 \\
$13959.999$ \dotfill &   $110.21$ &  5.10 \\
$13966.942$ \dotfill &    $81.34$ &  6.02 \\
$13975.925$ \dotfill &    $79.43$ &  5.48 \\
$13998.914$ \dotfill &   $104.92$ &  5.44 \\
$14001.894$ \dotfill &    $92.09$ &  5.13 \\
$14020.897$ \dotfill &    $92.71$ &  5.89 \\
$14035.837$ \dotfill &    $86.45$ &  6.82 \\
$14046.831$ \dotfill &    $90.77$ &  5.27 \\
$14059.809$ \dotfill &    $97.79$ &  6.12 \\
$14099.808$ \dotfill &    $90.56$ &  5.14 \\
$14130.751$ \dotfill &   $103.82$ &  6.21 \\
$14136.648$ \dotfill &    $93.97$ &  5.58 \\
$14150.684$ \dotfill &   $103.07$ &  8.31 \\
$14170.642$ \dotfill &    $72.56$ &  5.83 \\
$14304.984$ \dotfill &    $66.41$ &  7.94 \\
$14309.978$ \dotfill &    $71.15$ &  4.79 \\
$14311.997$ \dotfill &    $53.32$ &  4.67 \\
$14336.958$ \dotfill &    $61.70$ &  4.57 \\
$14401.815$ \dotfill &    $56.27$ &  6.58 \\
$14404.904$ \dotfill &    $55.17$ & 11.24 \\
$14405.854$ \dotfill &    $57.46$ &  5.66 \\
$14417.794$ \dotfill &    $30.49$ &  7.80 \\
$14421.776$ \dotfill &    $52.18$ &  5.22 \\
$14422.837$ \dotfill &    $56.29$ &  5.96 \\
$14423.791$ \dotfill &    $59.01$ &  4.68 \\
$14425.847$ \dotfill &    $64.05$ &  5.06 \\
$14426.798$ \dotfill &    $53.18$ &  3.70 \\
$14525.640$ \dotfill &    $20.61$ &  7.68 \\
$14548.644$ \dotfill &    $26.06$ &  5.19 \\
$14675.985$ \dotfill &   $-30.05$ &  6.84 \\
$14677.971$ \dotfill &   $-22.00$ &  6.32 \\
$14699.954$ \dotfill &   $-37.38$ & 11.05 \\
$14734.937$ \dotfill &   $-65.10$ &  5.39 \\
$14748.784$ \dotfill &   $-82.01$ &  6.90 \\
$14767.935$ \dotfill &   $-94.18$ &  5.89 \\
$14846.680$ \dotfill &    $38.50$ &  5.49 \\
$14858.657$ \dotfill &    $43.15$ &  7.74 \\
$14904.639$ \dotfill &    $78.14$ &  7.19 \\
\enddata
\end{deluxetable}

The radial velocities, measured in approximately
700 2\sA\ segments of each echelle spectrum, are averaged to
determine the radial velocity of a star in a given observation. 
Table \ref{vel16175} contains the velocities
and their associated uncertainties for 
\starB, from 2004 November to 2009 March.
We estimate those uncertainties to be the standard deviation from 
the mean of the 700 individual segment velocities.
Typical internal uncertainties for \starB\ are
$\sim$\,$6\unit{\ms}$.
The internal uncertainties in Table \ref{vel16175} are added
in quadrature to a $5\unit{\ms}$ error due to stellar
jitter (a typical figure for subgiant stars; see 
\citealt{fischer_2003} and \citealt{johnson_2007a}), 
and those combined values are used in the 
least-squares Keplerian analysis described
in \S\,\ref{keplerian_fit_b}.

\subsection{Stellar Properties}
\label{stellar_properties_b}

\starB\ is a G0 subgiant in the foreground
of galactic open cluster M34 (NGC\,1039), 
which is $440\unit{pc}$ farther away.
In our analysis we found several different literature
$V$-band magnitudes for \starB, so we adopted 
for our analysis here an average of the values from three 
sources:\ the \citet{tereshchenko_2001} 
spectrophotometric standards list, 
the \hip\ catalog \citep{hipparcos}, 
and the Sky2000 
catalog.\footnote{Vizier Online Data Catalog, 5109 \citep{sky2000}} 
Our adopted values were $V=\vmagB\pm0.03$ and 
$B-V=\bvB \pm 0.02$.

The \hip\ parallax distance for \starB\ is $\dB \pm \deB\unit{pc}$, 
which, when taken together with $V$ from the preceding paragraph, 
implies an absolute visual magnitude $M_V=\mvB$. 
Its $M_V$ places \starB~$1.3\unit{mag}$ above the 
\hip\ main sequence defined in \citet[a value that fulfills the 
selection criterion described in \S\,\ref{rv_measurements_b}]{wright_2005}.
\starB's high metallicity (measured in the next paragraph) 
may account for part of its distance above the main sequence,
but some of that $1.3\unit{mag}$ difference is likely due 
to its evolution, even though it may not yet have exhausted 
its core hydrogen.

\starB\ is metal rich, with \feh~$ = \mfeB \pm \feeB$, and has an
effective temperature $\mteff = \teffB \pm \teffeB\unit{K}$.
Metallicity and temperature come from our
LTE spectral synthesis analysis with
Spectroscopy Made Easy \citep[SME;][]{valenti_1996}. 
In general we follow the procedure outlined in \citet{valenti_2005}.
The LTE analysis also yielded 
surface gravity 
\logg~$ = \loggB \pm \loggeB$
and projected rotational velocity 
\vsini~$ = \vsiniB \pm \vsinieB\unit{\kms}$.
The uncertainties of the SME-derived stellar properties 
(\feh, \teff, \logg, and \vsini) were determined 
following the procedure described in
\citet{johnson_2008} and \citet{valenti_2005}.
The full set of stellar properties and their uncertainties
(in parentheses) appear in Table \ref{table_stellar}. 

The stellar luminosity 
$L_{\star} = \lstarB \pm \lstareB\unit{\lsun}$ 
was calculated using the relation in 
\citet{valenti_2005}, based on $M_{V}$ and a 
bolometric correction from \citet{vandenberg_2003}. 
We follow the same procedure here, adopting a
bolometric correction of $\bcorrB\unit{mag}$.
The stellar radius 
$R_{\star} = \rstarB \pm \rstareB\unit{\rsun}$ 
was calculated with the Stefan-Boltzmann law, 
using $L_{\star}$ given and \teff\ from the 
LTE analysis.
We then calculated the stellar mass 
$M_{\star} = \mstarB \pm \mstareB\unit{\msun}$ 
and age $\ageB \pm 1.0\unit{Gyr}$ 
using the
\citet{girardi_2002} stellar interior models, based on
the \hip-derived $M_{V}$, the $B-V$ color, 
and the LTE-derived metallicity.
The uncertainties for these derived values are propagated
from the uncertainties on the input values.

The chromospheric activity parameters 
$S_{\rm HK}$ and $\log R'_{\rm HK}$, based on the 
\caii\ H and K emission features, are measured 
according to the procedure in
\citet{wright_2004}.
The full listing of stellar characteristics for 
\starB\ appears in \mbox{Table \ref{table_stellar}}.

\def\arraystretch{1.15}
\begin{deluxetable}{lcc}[!hbtp]
\tablecaption{Stellar Parameters.\label{table_stellar}}
\tablewidth{0pt}
\tablehead{
  \colhead{\makebox[3.5cm][c]{Parameter}} &
  \colhead{\starB\tablenotemark{a}}    &
  \colhead{\starA\tablenotemark{a}}    
}

\startdata
$V$~(mag) \dotfill       & \vmagB        & \vmagA        \\
$M_V$~(mag) \dotfill     & \mvB          & \mvA          \\
$B-V$~(mag) \dotfill     & \bvB          & \bvA          \\
Distance~(pc) \dotfill   & \dB~(\deB)    & \dA~(\deA)    \\
\feh \dotfill            & \feB~(0.06)   & \feA~(0.06)   \\
\teff~(K) \dotfill       & \teffB~(\teffeB)   & \teffA~(\teffeA)   \\
\vsini~(\kms) \dotfill   & \vsiniB~(\vsinieB) & \vsiniA~(\vsinieA) \\
\logg \dotfill           & \loggB~(\loggeB)  & \loggA~(\loggeA)   \\
$M_{\star}$~(\msun) \dotfill    & \mstarB~(\mstareB) & \mstarA~(\mstareA) \\
$R_{\star}$~(\rsun) \dotfill    & \rstarB~(\rstareB) & \rstarA~(\rstareA) \\
$L_{\star}$~(\lsun) \dotfill    & \lstarB~(\lstareB) & \lstarA~(\lstareA) \\
Bolometric Corr.~(mag) \dotfill & $\bcorrB$     & $\bcorrA$     \\
Age~(Gyr) \dotfill           & \ageB~(1.0)   & \ageA~(1.0)   \\
$S_{\rm HK}$ \dotfill        & \shkB         & \shkA         \\
$\log R'_{\rm HK}$ \dotfill  & \rhkB         & \rhkA         \\
\enddata
\tablenotetext{a}{Uncertainty given in parentheses.}
\end{deluxetable}

\subsection{Keplerian Fit}
\label{keplerian_fit_b}

We search for the best fitting Keplerian orbital solution to the 
radial velocity time series with a Levenberg-Marquardt 
least-squares minimization. 
Beginning with a periodogram to determine an initial guess for 
the orbital parameters, our Keplerian fitter searches a grid
of initial period guesses to determine the orbital solution.
We estimate the uncertainties in those orbital parameters with a 
bootstrap Monte Carlo method:\ we subtract the best-fit Keplerian 
from the measured velocities, scramble the residuals and add them
back to the original measurements, then obtain a new set of
orbital parameters. 
We repeat this process for 1000 trials and adopt 
the standard deviations of the parameters from all trials 
as the formal uncertainties. 
For a more in-depth description of the Keplerian fitting 
and uncertainty determination procedures, 
see, e.g., \citet{marcy_2005}.

For \starB, the best fitting Keplerian orbital solution has a period
of $P = \pB \pm \peB\unit{days}$ 
(or $\pyearsB \pm \pyearseB\unit{yr}$). 
It also has eccentricity $e = \eB \pm \eeB$ and velocity semi-amplitude 
$K = \kB \pm \keB\unit{\ms}$. 
The rms scatter of the data about the fit is $\rmsB\unit{\ms}$, 
and the reduced \chisq\ is \chisB. 
Based on our stellar mass estimate in \S\,\ref{stellar_properties_b}
of $M_{\star} = \mstarB\unit{\msun}$, the planet \starB\,b
has a semimajor axis $a = \arelB\unit{AU}$ and 
a minimum planet mass $\msini = \msiniB\unit{\mjup}$. 
The full list of orbital parameters appears in Table \ref{table_orbits},
with their associated uncertainties in parentheses.

\begin{figure}[!b]
\epsscale{1.15}
\plotone{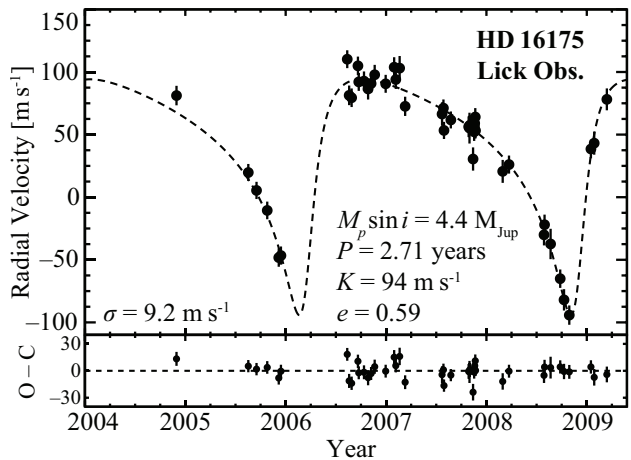}
\caption{Radial velocity time series for \starB, measured at Lick Observatory. The error bars represent a quadrature sum of the internal measurement error (listed in Table \ref{vel16175}) and $5\unit{\ms}$ stellar jitter (see \S\,\ref{keplerian_fit_b} for more detail). The {\em dashed line} represents the best Keplerian orbital fit to the data for \starB\,b. The parameters of that fit are printed in the plot inset. The residuals between the Keplerian fit and the data appear in the lower panel of the plot.
\label{orbitB}}
\end{figure}

The velocities from Table \ref{vel16175} are plotted in Figure \ref{orbitB}.
The error bars on the plot represent the quadrature addition of the
internal measurement errors from Table \ref{vel16175} and
$5\unit{\ms}$ of stellar jitter (see \S\,\ref{rv_measurements_b}). 
The best-fit orbital solution is plotted with a dashed line and 
the orbital parameters are printed in the plot.
The photometric stability and transit probability for 
\starB\,b are discussed in \S\,\ref{phot}.

\vspace{0.5cm}

\def\arraystretch{1.15}
\begin{deluxetable}{lcc}[!b]
\tablecaption{Orbital Parameters.\label{table_orbits}}
\tablewidth{0pt}
\tablehead{\colhead{\makebox[3cm][c]{Parameter}} &
  \colhead{\starB\,b\tablenotemark{a}}    &
  \colhead{\starA\,b\tablenotemark{a}}
}

\startdata
$P$ (d) \dotfill         & \pB~(\peB)              & \pA~(\peA)   \\
$T_p$\tablenotemark{b}~(JD) \dotfill & \tpB~(\tpeB) & \tpA~(\tpeA) \\
$e$ \dotfill              & \eB~(\eeB)              & \eA~(\eeA)   \\
$K$~(\ms) \dotfill        & \kB~(\keB)              & \kA~(\keA)   \\
$\omega$~(deg) \dotfill   & \omB~(\omeB)            & \omA~(\omeA) \\
$\msini$~(\mjup) \dotfill & \msiniB~(\msinieB)      & \msiniA~(\msinieA)  \\
$a$~(AU) \dotfill         & \arelB~(\areleB)        & \arelA~(\areleA)  \\
Fit RMS~(\ms) \dotfill    & \rmsB                   & \rmsA        \\
Jitter~(\ms) \dotfill     & 5.0                     & 5.0          \\
Reduced-\chisq \dotfill   & \chisB                  & \chisA       \\
$N_{\rm obs}$ \dotfill    & \nobsB                  & \nobsA       \\
\enddata
\tablenotetext{a}{Uncertainty given in parentheses.}
\tablenotetext{b}{Time of periastron passage.}
\end{deluxetable}

\vspace{0.5cm}

\section{\starA}    
\label{star_a}
\subsection{Observations and Radial Velocity Measurements}
\label{rv_measurements_a}

Doppler monitoring of \starA\ (HIP\,54195, BD\,$-$09\,3201) began 
at Keck Observatory in 2004 January as part of the N2K 
consortium observing program. 
N2K targeted 2000 stars
within $110\unit{pc}$ with the goal of detecting high transit
probability hot Jupiters.
The stars were chosen to have $0.4 < B-V < 1.2\unit{mag}$ and 
$V$ brighter than $10.5\unit{mag}$.
The targets were also chosen to be metal rich, with 
\feh\ of $+0.1$ or higher, as
determined photometrically \citep{ammons_2006} and with
low-resolution spectroscopy \citep{robinson_2006}.
Observations each star were collected 
in a high-cadence time series intended to detect
short-period planets. 
For a full description of the N2K consortium 
and its goals, see \citet{fischer_2005n2k}.
The main phase of the N2K program has finished, but 
several stars that showed significant velocity scatter have been 
monitored further. 
A number of longer period planets
have been detected around these stars, 
including the planet orbiting 
\starA\ announced here.

\starA\ was observed with the HIRES echelle spectrograph 
\citep{vogt_1994} on Keck I.
The spectra have $R \approx 70\,000$ at 
$\lambda = 5500$\sA. 
Typical exposure times were 2 minutes, yielding
SNR $\sim 250$ at 5000\sA.
The iodine cell procedure for measuring radial velocities in 
Keck spectra is the same as for Lick spectra, described in 
\S\,\ref{rv_measurements_b}.
As with the \lsps, we initially observed the N2K stars relative
to a morphed synthetic template spectrum (see 
\S\,\ref{rv_measurements_b}), 
taking a traditional template exposure once a planet candidate emerged.

The full list of measured velocities and their associated 
internal uncertainties (estimated as described in 
\S\,\ref{rv_measurements_b}) appears as Table \ref{vel96167}.
The observations span the period from 2004 January to 
2009 January. 
Typical uncertainties are $\sim$\,$1\unit{\ms}$,
and a jitter uncertainty of $5\unit{\ms}$ 
\citep{fischer_2003,johnson_2007a}
is added in quadrature before the least-squares
Keplerian fit is carried out (\S\,\ref{keplerian_fit_a}).

\subsection{Stellar Properties}
\label{stellar_properties_a}

\starA\ is a G5 subgiant star. 
In contrast to \starB, literature $V$-band magnitudes 
for \starA\ were consistent with \hip,
so we adopt the \hip\ values here: 
$V = \vmagA$ and $B-V = \bvA$.

\starA\ has a \hip\ parallax-based distance of 
$\dA \pm \deA\unit{pc}$.
The parallax distance and $V$-magnitude, when taken together, 
yield $M_V = \mvA$, placing it 
$1.5\unit{mag}$ above the main sequence as defined by 
the \citet{wright_2005} polynomial. 
\starA's high metallicity (measured in next paragraph)
may account for part of its distance above the main sequence,
but its $M_V$ places it far enough along the evolutionary
track for 1.3-\msun\ stars that it has likely exhausted
its core hydrogen and evolved into a subgiant.

Fitting with its inclusion in the N2K stellar sample, \starA\ is
metal rich, with \feh~$ = \mfeA \pm \feeA$, and has
a near-solar temperature $\mteff = \teffA \pm \teffeA\unit{K}$.
It has \logg~$= \loggA \pm \loggeA$ and
\vsini~$= \vsiniA \pm \vsinieA\unit{\kms}$.
Its mass is $\mstarA \pm \mstareA\unit{\msun}$ and its age is
$\ageA \pm 1.0\unit{Gyr}$; it has a luminosity $L_{\star}$ of
$\lstarA \pm \lstareA\unit{\lsun}$
and radius $R_{\star}$ of 
$\rstarA \pm \rstareA\unit{\rsun}$.
Its \caii\ H and K chromospheric activity is relatively quiet,
with $S_{\rm HK} = \shkA$ and $\log{R'_{HK}} = $~\rhkA.
As with \starB, the \feh, \teff, \logg, and  \vsini\ come
from an LTE spectral synthesis using SME.
The procedures for determining
$L_{\star}$, $R_{\star}$, $M_{\star}$, age, $S_{\rm HK}$,
and $\log{R'_{HK}}$ and their associated uncertainties are 
identical to those described in \S\,\ref{stellar_properties_b}.
The full set of stellar parameters is listed in
Table \ref{table_stellar}.

\subsection{Keplerian Fit}
\label{keplerian_fit_a}

The procedure for determining the best fitting Keplerian orbital 
solution for the velocities listed in Table \ref{vel96167} is 
the same as described in \S\,\ref{keplerian_fit_b}. 
The best fitting orbital solution has 
$P = \pA \pm \peA\unit{days}$ (or 
$\pyearsA \pm \pyearseA\unit{yr}$).
The eccentricity $e$ is $\eA \pm \eeA$ and the velocity 
semi-amplitude $K$ is $\kA \pm \keA\unit{\ms}$.
The residuals from the best fit have an rms scatter of 
$\rmsA\unit{\ms}$; the fit has a reduced \chisq\ of
\chisA.
Based on the LTE-determined mass estimate 
(\S\,\ref{stellar_properties_a}) of 
$M_{\star} = \mstarA\unit{\msun}$, the planet
\starA\,b has semi-major axis $a = \arelA$ and 
minimum mass $\msini = \msiniA\unit{\mjup}$.

\begin{figure}[!htp]
\epsscale{1.15}
\plotone{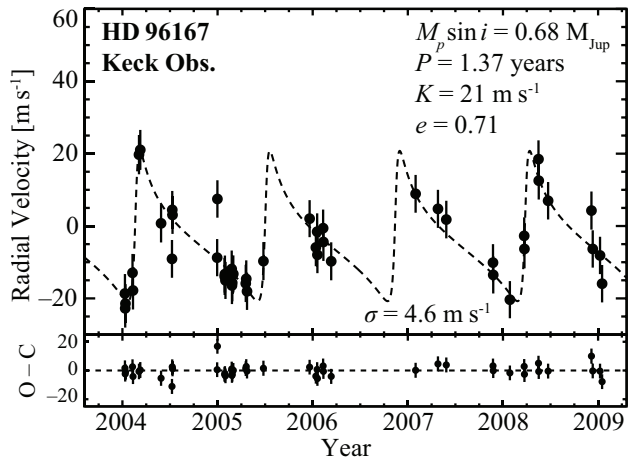}
\caption{Radial velocity time series for \starA, measured at Keck Observatory. The error bars represent a quadrature sum of the internal measurement error (listed in Table \ref{vel96167}) and $5\unit{\ms}$ stellar jitter (see \S\,\ref{keplerian_fit_a} for more detail). The {\em dashed line} represents the best Keplerian orbital fit to the data for \starA\,b. The parameters of that fit are printed in the plot inset. The residuals between the Keplerian fit and the data appear in the lower panel of the plot.
\label{orbitA}}
\end{figure}

The \starA\ radial velocities listed in Table \ref{vel96167} are plotted
in Figure \ref{orbitA}. As with \starB, the error bars represent
the quadrature sum of the internal measurement uncertainties from 
Table \ref{vel96167} and stellar jitter of 
$5\unit{\ms}$. 
The best-fit orbital solution appears as the dashed line in Figure 
\ref{orbitA}. 
The photometric stability and transit probability for 
\starA\,b are discussed in \S\,\ref{phot}.

\def\arraystretch{1.100}
\begin{deluxetable}{lrr}[!htp]
\tablecaption{Radial Velocities for HD\,96167.\label{vel96167}}
\tablewidth{0pt}
\tablehead{
 &
\colhead{Radial Velocity} &
\colhead{Uncertainty} \\
\colhead{\makebox[3cm][c]{${\rm JD}-{\rm 2,440,000}$}} &
\colhead{(\ms)} &
\colhead{(\ms)} 
}
\startdata
$13015.110$ \dotfill &   $-18.65$ &  1.99 \\
$13016.127$ \dotfill &   $-22.75$ &  1.82 \\
$13017.114$ \dotfill &   $-21.43$ &  1.98 \\
$13044.149$ \dotfill &   $-12.93$ &  1.48 \\
$13046.017$ \dotfill &   $-17.81$ &  1.62 \\
$13069.005$ \dotfill &    $19.73$ &  2.12 \\
$13073.938$ \dotfill &    $21.05$ &  2.32 \\
$13153.798$ \dotfill &     $0.74$ &  1.70 \\
$13195.752$ \dotfill &    $-9.08$ &  1.51 \\
$13196.783$ \dotfill &     $4.47$ &  1.53 \\
$13197.766$ \dotfill &     $3.06$ &  1.56 \\
$13369.127$ \dotfill &    $-8.74$ &  1.18 \\
$13370.114$ \dotfill &     $7.48$ &  1.33 \\
$13398.012$ \dotfill &   $-13.93$ &  1.14 \\
$13398.943$ \dotfill &   $-13.28$ &  1.11 \\
$13400.962$ \dotfill &   $-15.04$ &  1.10 \\
$13424.993$ \dotfill &   $-11.88$ &  1.03 \\
$13425.941$ \dotfill &   $-16.44$ &  1.16 \\
$13426.978$ \dotfill &   $-15.67$ &  1.20 \\
$13428.094$ \dotfill &   $-13.57$ &  1.14 \\
$13431.095$ \dotfill &   $-13.24$ &  1.10 \\
$13478.938$ \dotfill &   $-15.86$ &  1.20 \\
$13480.838$ \dotfill &   $-14.61$ &  1.11 \\
$13483.769$ \dotfill &   $-18.05$ &  1.16 \\
$13546.752$ \dotfill &    $-9.70$ &  1.62 \\
$13724.121$ \dotfill &     $2.03$ &  1.09 \\
$13748.102$ \dotfill &    $-5.92$ &  1.37 \\
$13753.044$ \dotfill &    $-1.56$ &  0.82 \\
$13754.051$ \dotfill &    $-7.94$ &  0.94 \\
$13775.991$ \dotfill &    $-0.60$ &  1.24 \\
$13776.990$ \dotfill &    $-4.41$ &  1.58 \\
$13806.985$ \dotfill &    $-9.75$ &  1.60 \\
$14131.028$ \dotfill &     $8.90$ &  1.50 \\
$14216.853$ \dotfill &     $4.74$ &  1.63 \\
$14248.838$ \dotfill &     $1.79$ &  1.41 \\
$14428.146$ \dotfill &   $-10.12$ &  1.05 \\
$14429.120$ \dotfill &   $-13.51$ &  1.15 \\
$14493.008$ \dotfill &   $-20.36$ &  1.17 \\
$14547.913$ \dotfill &    $-2.72$ &  1.10 \\
$14548.850$ \dotfill &    $-6.34$ &  1.96 \\
$14602.805$ \dotfill &    $18.48$ &  1.40 \\
$14603.795$ \dotfill &    $12.53$ &  1.46 \\
$14639.781$ \dotfill &     $6.98$ &  1.34 \\
$14806.156$ \dotfill &     $4.27$ &  1.66 \\
$14811.120$ \dotfill &    $-6.34$ &  1.25 \\
$14839.123$ \dotfill &    $-8.13$ &  1.28 \\
$14847.065$ \dotfill &   $-15.94$ &  1.31 \\
\enddata
\end{deluxetable}

\vspace{0.4cm}

\section{Photometric Measurements of \starB\ and \starA}  
\label{phot}
We have obtained three years of high-precision differential 
photometry of both \starB\ and \starA\ with the T12 \mbox{0.8 m} 
automatic photometric telescope (APT) at Fairborn Observatory.  
The APT detects short-term, low-amplitude brightness 
variability in solar-type stars due to rotational 
modulation in the visibility of photospheric starspots 
\citep[e.g.,][]{hfh95}, as well as longer-term variations 
associated with stellar magnetic cycles \citep{h99}.  
We obtain photometric observations to establish whether observed 
radial velocity variations in a star are due to reflex motion caused by a 
planetary companion or due to the effects of stellar activity 
\citep[e.g.,][]{qhs01,psc04}.  Photometric observations can also lead to
the detection of planetary transits and the direct determination of planetary
radii, as in \citet{wht08}.

We acquired 181 observations of \starB\ between 2006 December and 
2009 February and 319 observations of \starA\ between 2004 November 
and 2007 May.  
The T12 APT uses two temperature-stabilized EMI~9124QB photomultiplier 
tubes to measure photon count rates simultaneously through Str\"omgren $b$ 
and $y$ filters. 
On a given night, the telescope observes each target star 
and three nearby comparison stars, along with measures of 
the dark count rate and sky brightness in the vicinity of each star.  
Designating the comparison stars as A, B, and C, and the target star 
as T, the observing sequence is as follows:\ 
dark, A, B, C, T, A, sky$_{\rm A}$, B, sky$_{\rm B}$, C, sky$_{\rm C}$, 
T, sky$_{\rm T}$, A, B, C, T. 
For \starB, the comparison stars A, B, and C 
were HD\,16176, HD\,14095, and HD\,14064, respectively; for \starA, the
comparison stars were HD\,96220, HD\,95938, and HD\,94206. 
A diaphragm
size of $45\arcsec$ and an integration time of 20 seconds were used
for all integrations.  

\def\arraystretch{1.15}
\begin{deluxetable*}{rccrccc}[!htp]
\footnotesize
\tablewidth{0pt}
\tablecaption{Summary of the Ensemble Photometric Observations\label{phottable}}
\tablehead{
\colhead{} & \colhead{} & \colhead{Date Range} & \colhead{} &
\colhead{Yearly Mean} & \colhead{$\sigma$} & \colhead{$\sigma_{mean}$} \\
\colhead{\makebox[3cm][c]{Star}} & \colhead{APT} & \colhead{(JD$ - $2,440,000)} &
\colhead{$N_{obs}$} & \colhead{(mag)} & \colhead{(mag)} & \colhead{(mag)} \\
\colhead{(1)} & \colhead{(2)} & \colhead{(3)} & \colhead{(4)} &
\colhead{(5)} & \colhead{(6)} & \colhead{(7)}
}
\startdata
\starB \dotfill & T12 & 14080--14172 &  \makebox[1cm][r]{47} & 
\makebox[1.5cm][c]{0.41236} & 
\makebox[1.5cm][c]{0.00123} & \makebox[1.5cm][c]{0.00018} \\
                &     & 14370--14533 &  86 & 0.41232 & 0.00160 & 0.00017 \\
                &     & 14730--14881 &  48 & 0.41079 & 0.00124 & 0.00018 \\
\starA \dotfill & T12 & 13330--13511 & 130 & 0.98318 & 0.00136 & 0.00012 \\
                &     & 13690--13886 & 124 & 0.98380 & 0.00144 & 0.00013 \\
                &     & 14058--14255 &  65 & 0.98287 & 0.00107 & 0.00013 \\
\enddata
\end{deluxetable*}

The measurements in each sequence were reduced to form three 
independent measures of the six differential magnitudes 
$\mT-\mA$, $\mT-\mB$, $\mT-\mC$, $\mC-\mA$, $\mC-\mB$, and 
$\mB-\mA$.  
These differential magnitudes were corrected for extinction and 
transformed to the standard Str\"omgren photometric system. To increase
the S/N, the data from the $b$ and $y$ passbands
were averaged to create ``$(b+y)/2$'' magnitudes. After passing quality 
control tests, the three independent measures of each differential magnitude 
were combined to give a single mean data point per complete sequence for each 
of the six differential magnitudes. 

\begin{figure}[!hbp]
\epsscale{1.15}
\plotone{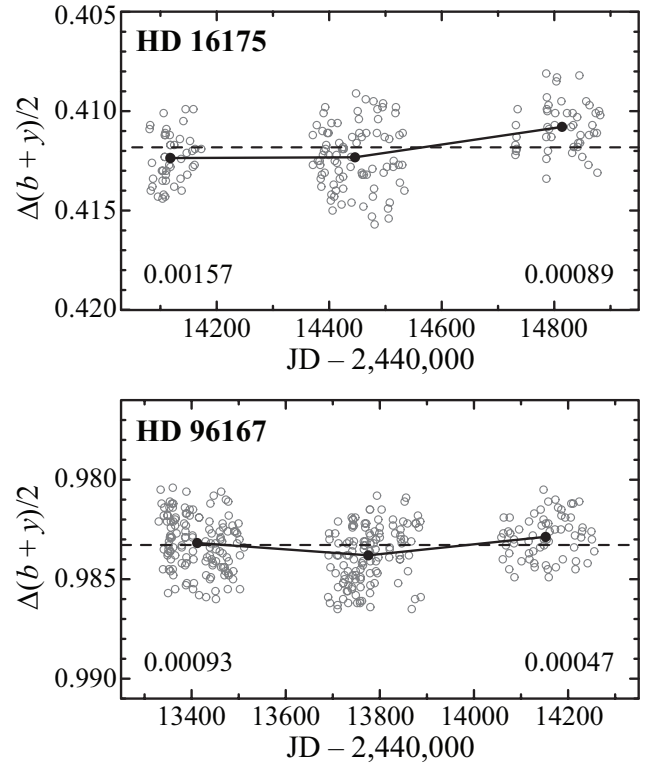}
\caption{Nightly mean differential magnitudes of both stars ({\em open circles}). The yearly means are plotted as {\em filled circles} connected with {\em straight line} segments. The uncertainties of the yearly means, as given in column 7 of Table \ref{phottable}, approximately match the size of the plotted symbols.
The {\em dashed line} in each panel represents the mean of the three yearly means.  The total range in magnitudes of the three yearly means is given in the lower left corner of each panel.  The standard deviation of the yearly means from the mean of the means ({\em dashed lines}) is given in the lower right corner.
\label{photfig}}
\end{figure}

To improve our measurement precision still further, we averaged the three
mean $\mT-\mA$, $\mT-\mB$, and $\mT-\mC$ differential magnitudes from 
each sequence
into a single value representing the difference in brightness between 
the program star and the mean of the three comparison stars 
$\mT-(\mA+\mB+\mC)/3$, 
which we refer to as the ensemble mean.  
This helps to average out any 
subtle brightness variations in the three comparison stars.  
The ensemble means for the three observing seasons of both 
stars are summarized in Table \ref{phottable} and plotted as
open circles in Figure \ref{photfig}.

Column 5 gives the yearly mean $\mT-(\mA+\mB+\mC)/3$ 
differential magnitudes in 
$(b+y)/2$ for the three observing seasons.  Column 6 gives the standard 
deviations, $\sigma$, of these ensemble differential magnitudes from the 
yearly mean and so provide a measure of night-to-night brightness 
variations.  Typical standard deviations for constant stars fall in the range 
0.0012--0.0017 mag for this telescope.  All six of the standard deviations 
in Table \ref{phottable} fall within this range.  Periodogram analyses of the three 
observing seasons for each star found no significant periodicity in the 
range of 1--100 days.  Both \starB\ and \starA\ are constant 
from night to night to the limit of our measurement precision; stellar 
activity due to starspots should have no significant effect on the measured 
radial velocities.

The standard deviations of the yearly means are given in column 7, computed 
from the $\sigma$ in column 6 divided by the square root of the number of 
observations in column 4; these values provide an estimate of the precision 
of each yearly mean.  
The observed yearly means are plotted in Figure \ref{photfig} as 
filled circles connected with solid line segments.  
For both \starB\ and \starA, the observed scatter in the yearly means 
is somewhat higher than the estimated precision of the means, by 
factors of 5.0 and 3.7, respectively, implying that there are 
slight year-to-year variations in the brightness of 
the two stars.  
\citet{h99} demonstrates that subtle brightness variability 
can be measured for solar-type stars and also demonstrates 
observationally that yearly means of constant stars can be measured to
0.0002 mag with the APTs.  
\citet{wmb08} includes another example of this measurement precision 
for comparison stars.  
Thus, the long-term variation over a range of 0.0016 mag in 
\starB\ is likely to be real.  
The range for \starA\ is less than a millimagnitude over the 
three observing seasons and so is harder to establish as real 
without a longer time series.
As subgiants with very low chromospheric activity levels 
($S_{\rm HK}$ and $\log{R'_{HK}}$ in Table \ref{table_stellar}), 
both stars are expected 
to have little brightness variation.  
Given the moderately long orbital periods and the current 
precision of the orbits, a transit search is premature.

\section{Summary}
\label{summary}
We present here two Jovian-mass planets in relatively long 
period, eccentric
orbits around intermediate-mass stars with supersolar
metallicity.
\starB\ is a $\mstarB\textrm{-\msun}$ G0 subgiant
from the \lsps\ at Lick Observatory,
orbited by a planet with a minimum mass of
$\msiniB\unit{\mjup}$ in a 
\pyearsB\ yr, eccentric ($e=\eB$) orbit.
\starA\ is a $\mstarA\textrm{-\msun}$ G5 subgiant
from the N2K program at Keck Observatory,
orbited by a planet with a minimum mass of
$\msiniA\unit{\mjup}$ in a
\pyearsA\ yr, eccentric ($e=\eA$) orbit.
The relatively high metallicities and masses of
the stars in question make these planets
particularly interesting additions to
the exoplanet menagerie. 

Approximately half of the planets orbiting subgiants 
detected to date are in orbits longer than a year. 
\starB\,b and \starA\,b are additions to that category, 
with periods of $\pyearsB \pm \pyearseB$ and
$\pyearsA \pm \pyearseA\unit{years}$, respectively.
Of the 267 known Doppler exoplanets with well constrained
orbits within $200\unit{pc}$, 23 are in orbits more eccentric 
than \starB\,b ($e = \eB$) and only 12 are in orbits 
more eccentric than \starA\,b ($e=\eA$).
Only eight of the known Doppler planets orbit stars more
metal rich than \starB\ (\feh = \feB), while only 23 
orbit stars more metal rich than \starA\ (\feh = \feA). 
The planets fall into the upper
reaches of three emerging trends in exoplanet 
distribution (eccentricity, host star mass, and 
metallicity), thereby contributing to our
understanding of the nature of exoplanets and
of planet formation.

\acknowledgements
We are indebted to the many observers who collected spectra 
of these stars. 
For \starB, we thank the observers
Kelsey Clubb, 
Julia Kregenow, 
Joshua Peek, 
Karin Sandstrom, 
and Julien Spronck.
For \starA, we thank the observers
Gaspar Bakos, 
R.~Paul Butler, 
Chris McCarthy, 
Guillermo Torres,
Steven S.~Vogt, 
and Joshua Winn.
We gratefully acknowledge the dedication and support of 
the Lick and Keck Observatory staffs, in particular 
Tony Misch for support with the Hamilton Spectrograph
and Grant Hill and Scott Dahm for support with HIRES.
JAJ is an NSF Astronomy and Astrophysics Postdoctoral Fellow with
support from the NSF grant AST-0702821.
DAF acknowledges research support from NASA grant NNX08AF42G.
GWH acknowledges that Automated Astronomy at 
Tennessee State University has been supported by NASA and NSF 
as well as Tennessee State University and the State of 
Tennessee through its Centers of Excellence program.
We thank the NASA Exoplanet Science Institute (NExScI) for 
support through the KPDA program. We thank the NASA, NOAO, and 
UCO/Lick telescope assignment committees for allocations of 
telescope time.  
The authors extend thanks to those of Hawaiian ancestry on whose 
sacred mountain of Mauna Kea we are privileged to be guests.  
Without their kind hospitality, the Keck observations presented 
here would not have been possible. This research has made use of 
the SIMBAD database, operated at CDS, Strasbourg, France, and of 
NASA's Astrophysics Data System Bibliographic Services.



\bibliographystyle{apj}

\end{document}